\documentclass{emulateapj}

\slugcomment{To appear in ApJL., ???}

\shorttitle{SBS~1543+593}
\shortauthors{Schulte-Ladbeck et al.}

\begin{document}


\title{Emission-line spectroscopy of Damped Lyman Alpha Systems: The case of SBS~1543+593/HS~1543+5921}

\author{Regina E. Schulte-Ladbeck\altaffilmark{1},     
        Brigitte K\"onig\altaffilmark{1}, 
        Christopher J. Miller\altaffilmark{2},
        Andrew M. Hopkins\altaffilmark{1},
        Igor O. Drozdovsky\altaffilmark{3},
        David A. Turnshek\altaffilmark{1}, 
	and Ulrich Hopp\altaffilmark{4} 
}

\email{rsl@phyast.pitt.edu, bkoenig@phyast.pitt.edu,
cmiller@noao.edu, ahopkins@anu.phyast.pitt.edu, dio@ipac.calthech.edu,
turnshek@quasar.phyast.pitt.edu, hopp@usm.uni-muenchen.de}
\altaffiltext{1}{Physics \& Astronomy Department, University of
Pittsburgh, Pittsburgh, PA 15260, USA} \altaffiltext{2}{Cerro Tololo
Inter-American Observatory, Chile} \altaffiltext{3}{Spitzer Science
Center, California Institute of Technology, Pasadena, CA 91125, USA}
\altaffiltext{4}{Universit\"atssternwarte M\"unchen, Scheinerstr. 1,
D-81679 M\"unchen, Germany}

\begin{abstract}
We report HST/STIS spectroscopy and Gemini/GMOS-N imaging of the
Damped Lyman Alpha (DLA) system toward HS~1543+5921 caused by the host
star-forming galaxy (SFG) SBS~1543+593. The Gemini image shows new
morphological details of this well resolved DLA galaxy. In combination
with previous optical spectra, the new UV spectra enable us to compare
for the first time, ionized and neutral gas-phase alpha-element abundances 
derived from emission- and absorption-line spectroscopy, in a bona fide DLA
galaxy. The abundances we determine using emission-line diagnostics agree
with those from absorption-line diagnostics. We present our results on
a metallicity versus redshift diagram that combines local HII regions
and SFGs with high-redshift DLAs, and discuss implications for the chemical
evolution of galaxies.
\end{abstract}


\section{Introduction}

Damped Lyman Alpha systems on the sightlines to background quasars
exhibit high neutral hydrogen column densities and are thought to
arise in early disk galaxies capable to sustain star formation
\citep{wolfe1990}.  But over the last decade, only about a dozen DLA
{\it~galaxies} have been identified and imaged \cite[e.g.][]{turn2002,
chen2005}.

The chance projection of the galaxy SBS~1543+593 (z=0.0096) and the
background QSO HS~1543+5921 (z=0.807) 2.4'' away from the center of
the galaxy (Fig.~\ref{fig:sbs}) was discovered by \citet{reimers1998}.
\citet{bowen2001} and \citet{bowen2001b} noticed that the galaxy gives
rise to a DLA system in the spectrum of the QSO. \citet{sl2004a}
studied SBS~1543+593 using HST imaging and ground based
spectroscopy. They classified it an Sm dwarf galaxy with properties
that are entirely in line with those of other local dwarf
galaxies. Several of 33 HST-discovered HII regions were analyzed with
emission-line diagnostics. For the brightest region (\#5) all emission
lines necessary to derive the O and N abundances were detected.

SBS~1543+593 offers an excellent opportunity to directly compare
element abundances inferred from cool interstellar gas (DLAs) and
ionized gas (SFGs). None of the previously imaged DLA galaxies
resolves to show individual HII regions. In none of them, does the
sightline to the QSO intercept the disk of the galaxy well within its
optical radius, let alone, close to its center, thus eliminating concerns
over metallicity gradients. SBS~1543+593 facilitates
an important inquiry into the behavior of galaxy metallicities with
redshift, because almost all available abundances at high redshifts
are determined from QSO absorption line studies and refer to iron-peak
elements such as Zn and Cr \citep{pettini2004}, whereas at low
redshifts, derivations of chemical abundances are based on nebular
emission-line studies in star-forming regions, and are primarily,
based on the alpha-capture element oxygen
\citep[e.g.][]{garnett2004}. \cite{chen2005} recently derived O
abundances from emission-lines in one DLA and one sub-DLA galaxy; the
QSOs' spectra allowed them to determine absorption-line abundances of
Fe. These data are not suited to the comparison we are interested in
here, because Fe is a highly depleted element; we do not know the
$\alpha$/Fe ratio; and their QSOs have impact parameters larger than
the optical disks of the galaxies. SBS~1543+593 is unique, since it
enables a direct comparison between the emission- and absorption-line
techniques, and allows alpha element abundances of a DLA and its host
galaxy to be contrasted in a low-redshift QSO-galaxy pair.

\section{Observations}

Two sets of HST/STIS observations of the QSO HS~1543+5921 were
performed (ID 9784, PI Bowen) using grating G140M with a
dispersion of 0.05\,\AA/pix and centered on 1222.0\,\AA~and on
1272.0\,\AA. We averaged the five exposures per configuration to one
single spectrum in order to increase the signal-to-noise ratio.

The spectrum centered on 1222.0\,\AA\ covers the NI(1199.5\,\AA,
1200.2\,\AA, 1200.7\,\AA) triplet. The continuum is very noisy; and
the lines are located in the blue wing of the Galactic Lyman $\alpha$
line. For each line, we measure a lower and upper limit to its EW, by
varying the continuum placement. This results in 6 EW values, each of
which has a large uncertainty. The range a EWs will be translated into
an abundance range below.

\begin{table}
\begin{center}
\caption{Column densities derived from SII lines.\label{tab:SII}}
\begin{tabular}{cccccc}
\\
\colrule\colrule
Rest-frame & observed & EW    & REW   & $f$     &   $N$ \\

[\AA] & [\AA] & [\AA] & [\AA] & [cm]    &   [$10^{15}{\rm cm}^{-2}$] \\
\colrule
1250.58   & 1262.54 & $0.061\pm0.023$ & 0.060 & 0.00545 & $0.80\pm0.30$ \\
1253.81   & 1265.85 & $0.173\pm0.034$ & 0.171 & 0.01088 & $1.13\pm0.23$ \\
1259.52   & 1271.61 & $0.243\pm0.050$ & 0.241 & 0.01624 & $1.06\pm0.21$ \\
\colrule
\end{tabular}
\end{center}
\end{table}

The spectrum centered on 1272.0\AA\ (see Fig.~\ref{fig:SII}) is used
to measure the individual lines of the SII triplet. The continuum
level of SII($\lambda_{\rm obs}= 1262.54$\,\AA) is difficult to
determine and therefore uncertain; we consider this measurement as a
lower limit. SII($\lambda_{\rm obs}=1265.85$\,\AA) is unblended and
the continuum level is well determined. SII($\lambda_{\rm
obs}=1271.61$\,\AA) is blended with a nearby SiII line. We used two
Gaussian profiles to deblend the two lines. For an overview of the
measured values see Tab.~\ref{tab:SII}.


\begin{figure}
\epsscale{1.10}
\plotone{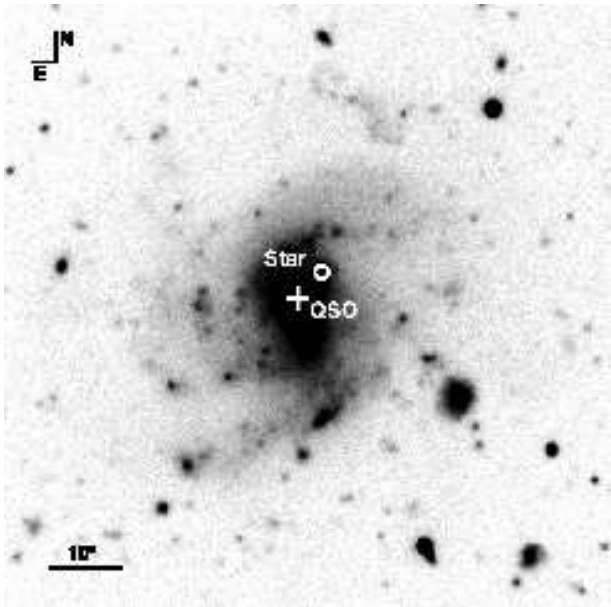}
\caption{GMOS-N observations of SBS~1543+593 using the r\_G0303
filter. To facilitate comparison with Fig.~1 of \citet{sl2004a}, we
marked the position of the QSO which is located in the background of
the galaxy, and a foreground star. Notice the extended spiral-arm
system of the galaxy that is not visible on the HST images of
\citet{sl2004a}.
  \label{fig:sbs}}
\vspace{-0.5cm}
\end{figure}

We performed deep optical imaging using GMOS-N (see
Fig.~\ref{fig:sbs}) equipped with the r\_G0303 filter, a broad band
filter (5620 - 6980\,\AA) with an effective wavelength of 6300\,\AA. A
total of six dithered 300\,sec exposures were taken. After pipeline
calibration the images were averaged to increase the signal-to-noise
ratio.

The HST image published by \citet{sl2004a} showed only the inner
saturated region of the new GMOS-N image. We now detect extended
spiral arms with associated HII regions. Note also the two features to
the NE which are only revealed in this deep image. The image suggests
a morphology which is still consistent with the previous
classification of SBS~1543+593 as an Sm galaxy.
Photometry of 193 regions within the image which are more extended
than the stellar point-spread function was performed using the IRAF
task DAOPHOT. Up to 140 could be HII regions, based on the fact that
they lie within the disk of the galaxy. Spectroscopy is needed to
confirm their nature and to measure their chemical abundances.

\section{Abundance analysis - neutral gas}
\begin{figure}
\plotone{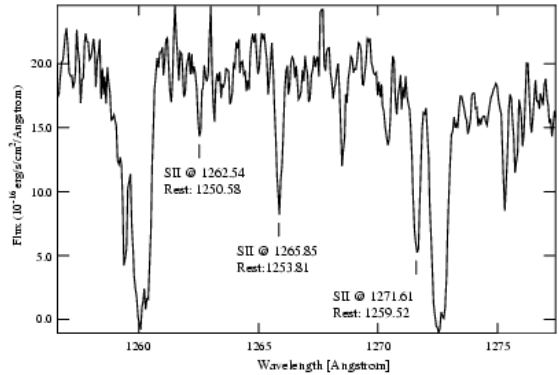}
\caption{HST/STIS spectrum centered on 1272.0\,\AA\ showing the SII
  triplet. The spectrum was smoothed with a boxcar of 3 pixel.
  \label{fig:SII}}
\vspace{-0.6cm}
\end{figure}

The abundances of sulfur and nitrogen are derived using the
measurements of the EWs and calculating the rest-equivalent width
($REW$) using the redshift of the galaxy. Together with
the oscillator strengths or $f$ values of \citet{morton1991} we derive
column densities ($N$) from eq.~8 of \citep{petitjean1998}.  This
expression is only applicable to optically thin absorption lines on
the linear part of the curve of growth.  Table~\ref{tab:SII} shows
that the equivalent widths of the three S lines used, are in the same
ratio as their $f$ values (within the errors), so assuming the
optically thin case appears to be valid. However, we must be concerned
that the spectral resolution is insufficient to discern if there are
any narrow saturated components contributing to the absorption
lines. If such components are present, we may have underestimated the
column density of S (but this would not invalidate the main result of
the paper).

In the following we will introduce a new nomenclature\footnote{${\rm
[Y/X]} = \log (Y/X) - \log (Y/X)_\odot$}. ${\rm [Y/X]_{\rm I}}$\
refers to the element abundance of element Y with respect to element X
in the neutral gas measured using absorption line diagnostics while
${\rm [Y/X]_{\rm II}}$\ refers to it measured in HII regions by
emission line diagnostics.

Each of the 6 nitrogen $REW$s is converted to a column
density. Considering the above-mentioned measurement uncertainties, we
give a range, ($0.12$ -- $1.80$)\,$10^{14}$\,cm$^{-2}$, rather than a
mean value for the column density. Consequently, we derive $-2.6
\lesssim {\rm [N/H]_I} \lesssim -2.0$, and adopting the solar nitrogen
abundance of $7.931\pm0.111$ from \cite{holweger2001}.

We use a solar sulfur abundance of 7.20 \citep{grevesse1998}. This is
consistent with \citet{prochaska2003b}. Note that \citet{holweger2001}
does not provide a sulfur abundance for the Sun. The REWs
($\lambda\lambda = 1254,\ 1259$) listed in Tab.~\ref{tab:SII}, were
used to calculate $N({\rm SII}) = (1.10\pm0.31)\ 10^{15}$\,cm$^{-2}$\
or ${\rm [SII/H]_I} = -0.50\pm0.33$.

The ionization energies for sulfur are SI: 10.36\,eV and for SII:
23.33\,eV. Abundance corrections in DLAs for ionization energies higher
than $h \nu > 13.6$\,eV are not necessary because these absorbers are
optically thick to such radiation. We therefore do not apply a
correction for SIII. For photons with energies $h\nu < 13.6$\,eV the
HI gas is transparent and thus $N({\rm SII})=N({\rm S})$. 

Assuming that DLAs arise in a neutral plus a mildly ionized region,
\citet{vladilo2001} model ionization correction terms for
sulfur. Depending on the ionization parameter U (the surface flux of
ionizing photons divided by the hydrogen particle density) the
correction would be between $0.15\geq \log(C{\rm [S/H]_I}) \geq
-0.20$\ for $\log N(H)= 20.34$ (their Fig.~9). We consider this range
of correction terms as an indication for the size of possible
systematic errors in our abundance determination.

Thus we derive a sulfur abundance of ${\rm [S/H]_I} = -0.50\pm0.33$,
or $0.29$\ solar.

\section{Abundance analysis - ionized gas}
We use [SII] fluxes from region \#5
\citep{sl2004a}. The rest-frame fluxes are
$F_{6717} = (0.61\pm0.09) 10^{-16}$\,erg\,s$^{-1}$\,cm$^{-2}$,
$F_{6731} = (0.46\pm0.09) 10^{-16}$\,erg\,s$^{-1}$\,cm$^{-2}$, $F_{\rm
H \beta} = (2.13\pm0.43) 10^{-16}$\,erg\,s$^{-1}$\,cm$^{-2}$. The
electron temperature is $T_{\rm e} = 11300^{+2100}_{-1500}$\,K; and
the electron density is $n_{\rm e} \approx 80$\,cm$^{-3}$.

We use IONIC \citep{shaw1995} to derive abundances from the line-flux
measurements. This results in ${\rm [SII/H]_{II}}=-1.25\pm0.30$\ for
the HII region. Errors are estimated by exploring the effects on the
outcome of the model when varying the input values with their errors.
As a consistency check, we also calculate the abundance of SII/H by
applying Eq.~11 of \citet{pagel1992}. This gives ${\rm
[SII/H]_{II}}=-1.31\pm0.45$. The errors are calculated by error
propagation. In the further text we will refer to the value derived
by using IONIC.

Note that the spectra of \citet{sl2004a} do not cover the strong
[SIII] lines at 9069\,\AA, 9532\,\AA. The weak [SIII] line at
6312.06\,\AA\ was covered by the WHT ISIS R600R/TEK4 spectrum
observed on 25 April 2000. We measure a rest-frame flux of $F_{\rm
6312} = (0.11\pm0.09)\ 10^{-16}$\,erg\,s$^{-1}$\,cm$^{-2}$\ in HII
region \#5. This leads to ${\rm [SIII/H]_{II}}=-0.32\pm0.30$. To
estimate the error we again explored the outcome of IONIC by varying
the input values with their errors. Thus we derive ${\rm [S/H]}_{\rm
II} = {\rm [SII+SIII/H]}_{\rm II} = -0.27\pm0.30$\ for the total
sulfur abundance.
We calculate ${\rm[SIII/H]}_{\rm II}=-0.36\pm0.40$ using eq.~12 of
\citet{pagel1992} as a consistency check, thus ${\rm
[S/H]_{II}}=-0.31\pm0.30$.  As another check we apply a correction
factor from photoionization models to the measured SII abundance to
account for SIII in the HII region. \citet{garnett1989} intensively
studied sulfur in extragalactic HII regions. Typically one observes
two major ionization states: SII and SIII.  \citet{sl2004a} measured
$({\rm OII}/{\rm OI})_{\rm II} = 0.41$. With this oxygen ratio we can
correct for the unseen SIII. Using Fig.~4 from \citet{garnett1989}
leads to a correction factor of $\log ({\rm SII}/{\rm SIII})_{\rm
II}=-1.0\pm0.2$; and we derive ${\rm [S/H]_{II}} = -0.21\pm
0.30$. Whereas SIV becomes important in highly ionized nebulae, or if
$({\rm OII}/{\rm OI})_{\rm II} < 0.3$, the measured $({\rm OII}/{\rm
OI})_{\rm II}$ ratio suggests that this ion does not make an important
contribution in our case.

\section{Results}

\begin{figure}
\includegraphics[width=0.45\textwidth]{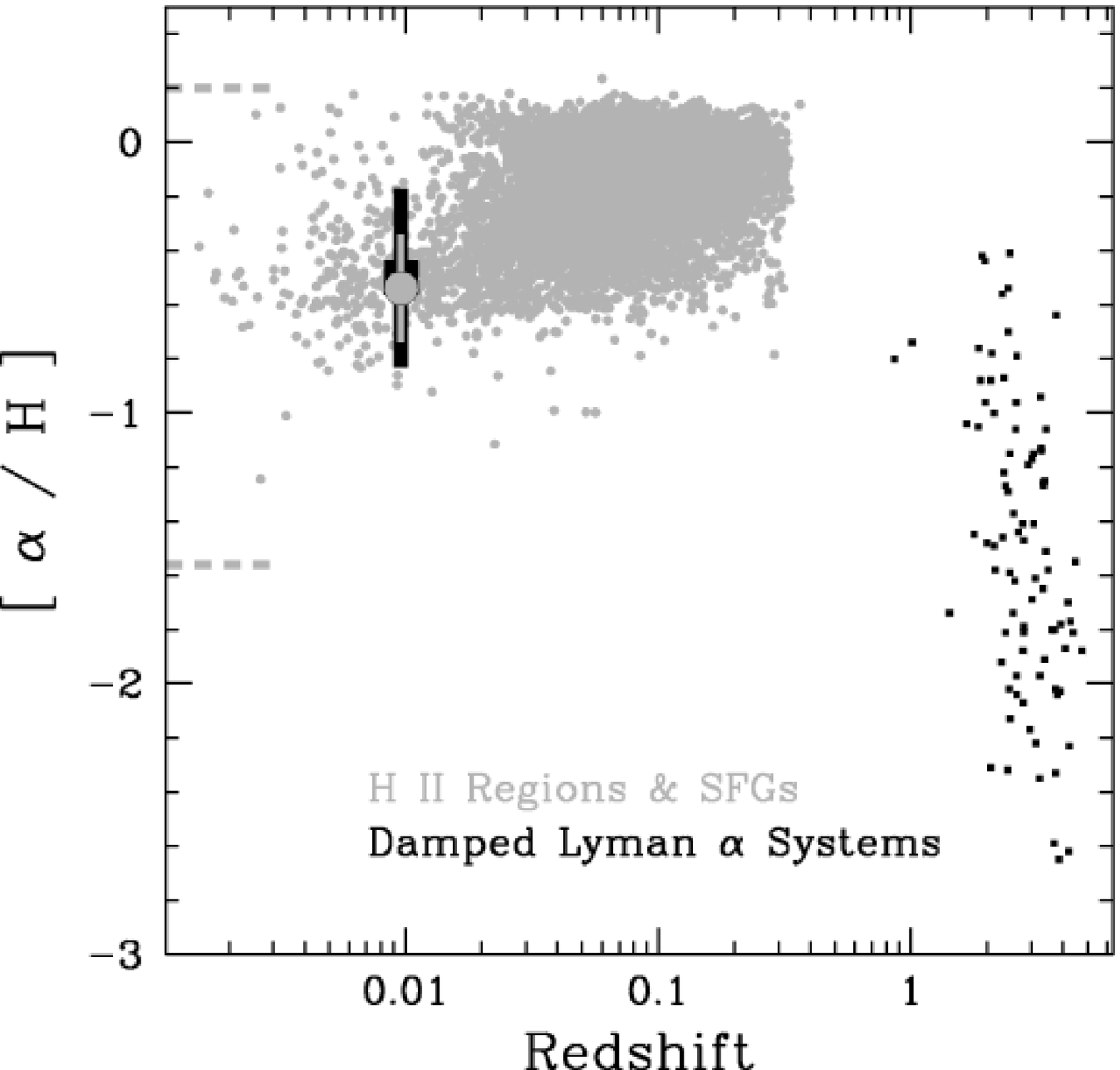}
\caption{Metallicity as a function of redshift. The dashed lines
  indicate the minimum and maximum local-universe values of ${\rm
  [O/H]}_{\rm II}$ (see text). The bright circles are
values of ${\rm [O/H]}_{\rm II}$\ for HII regions and SFGs from the
SDSS. The black squares are values of ${\rm [O,S,Si/H]}_{\rm I}$\ from
the compilation of \citet{prochaska2003d}. The neutral ([S/H]$_{\rm
I}$) and ionized ([O/H]$_{\rm II}$) gas phase abundances of
SBS~1543+593 are overlayed. Their agreement indicates that the
metallicities of SFGs and DLAs can be interpreted within the same
framework.
  \label{fig:sbsvsz}}
\vspace{-0.6cm}
\end{figure}
The measured sulfur column density in the neutral gas, $N({\rm SII}) =
1.10\ 10^{15}$\,cm$^{-2}$, is consistent with the prediction of
\citeauthor{sl2004a}, $N({\rm SII}) = 1.15\ 10^{15}$\,cm$^{-2}$.

They used three methods to determine ${\rm [O/H]}_{\rm II}
= -0.54\pm0.20$. HST/FOS data revealed ${\rm
[O/H]}_{\rm I}>-2.14$.  Here we find ${\rm [S/H]}_{\rm II} =
-0.27\pm0.30$, and ${\rm [S/H]_I} = -0.50\pm0.33$. Within the
errors these four values are the same.
\citeauthor{sl2004a} derived log ${\rm (N/O)}_{\rm II} =
-1.40^{+0.20}_{-0.30}$. Here, we calculate a range of $-2.0 \lesssim
\log {\rm(N/S)_I} \lesssim -0.8$. Again, we find agreement.

\section{Discussion}

Figure~\ref{fig:sbsvsz} offers a comparison between the metallicities
([O/H]$_{\rm II}$) of local SFGs and the metallicities ([O, S,
Si/H]$_{\rm I}$) of high redshift DLAs. Synthesized in massive stars,
O and Si are both made by alpha capture, while S originates from the
explosive burning of O \citep{Woos2002}. \citet{Nissen2002} recently
studied S in Galactic halo stars, and showed empirically that S
behaves like an alpha-capture element. Thus we use it as a proxy for
O. Note that the gas-phase Si abundances may be slightly
underestimated in DLAs since Si depletes onto dust grains.

In the local universe, accurate oxygen abundances are determined using
the direct, or $T_{\rm e}$ method. I~Zw~18 is the local galaxy with
the lowest oxygen abundance on record, $-1.56\pm0.01$
\citep{IT1999}. Even the analysis of a large sample of Sloan Digital
Sky Survey (SDSS) spectra using the $T_{\rm e}$ method
\citep{Kniazev2004} has not turned up a galaxy with a lower oxygen
abundance. Neither has a lower abundance been found in the outskirts
of local spirals \citep{Zee1998}. The local HII region with the
highest known oxygen abundance, $0.2\pm0.2$, resides in NGC~1232
\citep{Cast2002}. These minimum and maximum values are indicated in
Fig.~\ref{fig:sbsvsz} by dashed lines.

To illustrate the oxygen abundances at low redshifts, we plot on
Fig.~\ref{fig:sbsvsz}, oxygen abundances for local HII regions and
SFGs from the SDSS, using the selection criteria described in
\citet{sl2004b} and applying the O3N2 strong-line indices of
\citet{PP04} after adding 1.3\,\AA\ to the rest equivalent widths of
H$\alpha$\ and H$\beta$ to account for underlying stellar Balmer
absorption \citep{hop03}. The median error based on line fluxes is
0.05; and following Pettini \& Pagel, the systematic errors are 0.14
(their eq.~3) and 0.18 (their eq.~2). Our illustrative [O/H] values
for the redshift range from 0 to 0.3, span the range from $-1.2$ to
0.2, and represent well, local $T_{\rm e}$-based values. Note that our
median metallicities are systematically lower than those derived by
\citet{Tremonti} for SDSS SFGs; this is presumably due to different
methodologies. Adopting their [O/H] ratios would exacerbate the
difference between SFG and DLA metallicities.
  
For the DLAs, we plot on Fig.~\ref{fig:sbsvsz} the ${\rm [O/H]}_{\rm
I}$, ${\rm [S/H]}_{\rm I}$, and ${\rm [Si/H]}_{\rm I}$ ratios from the
compilation of \citet{prochaska2003d}. The median [$\alpha$/H] error in
this sample is 0.1.

Our main observations are the following: ${\rm [O,S,Si/H]}_{\rm I}$ of
high-redshift DLAs, go down to $-2.6$. Such low metallicities are
unobserved in the local universe!  DLA metallicities reach as high as
${\rm[O,S,Si/H]}_{\rm I}$ of $-0.4$, but never once do they attain
super-solar values, observed in local galaxies. Are these two
populations at all comparable?

We enter on Fig.~\ref{fig:sbsvsz} the values for SBS1543+593 as
follows. For the neutral-gas abundance, we show ${\rm [S/H]_I}$, and
for the ionized-gas abundance, we show ${\rm [O/H]}_{\rm II}$.  Here
is one DLA for which we can demonstrate that in principle, emission-
and absorption-line techniques give the same results when chemical
elements with similar nucleosynthetic origins are compared at similar
locations within a DLA galaxy.

There have long been discussions about whether or not emission- and
absorption-line diagnostics yield the same chemical abundances in SFGs
when one uses the internal star-forming regions rather than a QSO as
the background light source for the absorption experiment
\cite[e.g.][]{pettini1995, pettini2002, aloisi2003,
lecavelier2004}. There is as yet no agreement on the issue. A few
recent studies have addressed chemical abundances in (sub-)DLAs; but
these works do not compare the same elements in absorption and in
emission \cite[e.g.][]{christensen2005, ellison2005, chen2005}. The
case we have here doesn't suffer from any of these
uncertainties. Furthermore, in other DLA galaxies, the QSO intercepts
the galaxy's disks at much larger impact parameter, so internal
metallicity gradients have to be accounted for. Therefore, our's is
the cleanest comparison made to date between emission- and
absorption-line abundances in a DLA/SFG.

The agreement between the metallicity in the neutral and the ionized
gas phase of SBS~1543+593 validates in principle, the comparisons between 
the two types of objects, SFGs and DLAs, on the metallicity versus redshift
diagram. SBS~1543+593 thus makes the case that metal enrichment has taken place 
in the gas-rich, star-forming galaxy population between redshifts of 5 and 0.

In practice, galaxy metallicity gradients must be important for a
detailed comparison between SFG and DLA metallicities. DLAs are
merely defined by a neutral H column density threshold; yet we do not
know where a given DLA line is produced within a high-redshift
galaxy's disk.  Galaxy metallicity gradients have been investigated
for local disk galaxies; but it remains to be seen how they behave as
a function of redshift. First insights on this issue are provided by
the work of \cite{chen2005}.

\section{Conclusions}
We investigated O/H, S/H and N/O ratios in the ionized gas of the DLA
galaxy SBS~1543+593, and compared it with O/H, S/H, and N/S ratios in
its neutral gas. We find that the metallicities in the ionized and
neutral gas agree within the errors. The experiment allows us to
interpret DLA metallicities as an extension of SFG metallicities to
high redshift, suggesting that gas-rich galaxies had lower
metallicities at higher redshifts, when they were younger.

\acknowledgments

This paper is based in part on HST archival data. We thank David Bowen
for communicating his preliminary [S/H] results while the data were
still proprietary. RS-L and AMH gratefully acknowledge support awarded
by the Space Telescope Science Institute, which is operated by the
Association of Universities for Research in Astronomy, Inc., for NASA,
under contract NAS 5-26555.  Max Pettini and Sandhya Rao are thanked
for useful discussions of the paper. Please note that
\cite{prochaska2003d} contains a long list of original references
which we did not repeat here in the interest of brevity.  
We acknowledge the use of the SDSS archive.
The SDSS Web site is http://www.sdss.org/ and lists all partners.  
Based on
observations (GN-2004A-Q-82) obtained at the Gemini Observatory.
We thank Marcel Bergmann for assistance with the Gemini observations.
 






\end{document}